\documentclass[preprint,showpacs,preprintnumbers,amsmath,amssymb]{revtex4}

\usepackage{graphicx}
\usepackage{dcolumn}
\usepackage{bm}

\def \lsim {\,{\scriptscriptstyle{\stackrel{<}{\sim}}}\,}

\newcommand{\nn}{\noindent}
\newcommand{\beq}{\begin{equation}}
\newcommand{\eeq}{\end{equation}}

\begin{document}
\title{
Constraining New Forces in the Casimir Regime Using the
Isoelectronic Technique }
\author{R.~S.~Decca,${}^{1,*}$ D.~L\'{o}pez,${}^{2}$ H.~B.~Chan, ${}^{3}$
E.~Fischbach,${}^{4}$ D.~E.~Krause,${}^{5,4}$  and
C.~R.~Jamell${}^{1}$}

\affiliation{$^{1}$Department of Physics, Indiana
University-Purdue University Indianapolis, Indianapolis, Indiana
46202, USA \\
$^{2}$Bell Laboratories, Lucent Technologies, Murray Hill, New
Jersey 07974, USA \\
$^{3}$Department of Physics, University of Florida, Gainesville,
Florida 32611, USA \\
$^{4}$Department of Physics, Purdue University, West Lafayette,
Indiana
47907, USA \\
$^{5}$Physics Department, Wabash College, Crawfordsville, Indiana
47933, USA}

\today

\begin{abstract}
We report the first isoelectronic differential force measurements
between a Au-coated probe and two Au-coated films, made out of Au
and Ge. These measurements, performed at submicron separations
using soft microelectromechanical torsional oscillators, eliminate
the need for a detailed understanding of the probe-film Casimir
interaction. The observed differential signal is directly
converted into limits on the parameters $\alpha$ and $\lambda$
which characterize Yukawa-like deviations from Newtonian gravity.
We find $\alpha \lsim 10^{12}$ for $\lambda \sim 200$ nm, an
improvement of $\sim$ 10 over previous limits.
\end{abstract}

\pacs{03.70.+k, 12.20.Ds, 12.20.Fv, 42.50.Lc}

\maketitle

Although gravity was the first fundamental force to be understood,
the quest to unify it with the other fundamental forces has
remained elusive. One of the reasons is the apparent weakness of
the gravitational interaction at small separations. Consequently,
a significant number of experimental searches for new forces over
ultra-short distances has been performed
\cite{ours2,ours1,PLA,review,Long,Experimental1,Experimental2,Experimental3,Experimental4}.
They have been stimulated by at least three different---but
related---motivations:

(a) Some unification theories, incorporating $n$ compact extra
spatial dimensions with characteristic size $R(n)$, predict
deviations from Newtonian gravity over sub-mm scales
\cite{l18,Adelberger,l19}. Some extra-dimensional theories
characterize the deviations by a Yukawa-modified potential $V(r)$
\cite{l18,Adelberger,l19},

\beq V(r) = V_N (r) [1 + \alpha e^{-r/\lambda}]~,
   \label{1}
\eeq

\nn where $V_N(r)$ is the Newtonian gravitational potential for
two point masses separated by a distance $r \gg R(n)$, and
$\alpha$ and $\lambda \sim R(n)$ are constants.  Thus, the
extra-dimensional theories of Ref.~\cite{l18} provide a parameter
$R(n)$, which for $n > 1$ is naturally small, and an associated
constant $\alpha_n$ which is relatively poorly constrained.

(b) String theory and other extensions to the Standard Model
predict the existence of new light bosons such as  dilatons,
moduli, radions, cosmons, and bulk gauge bosons \cite{Adelberger}.
The exchange of these massive particles leads to corrections to
gravity as in Eq.~(\ref{1}), with $\alpha \gg 1$ and $\lambda$,
related to the boson mass $m$ by $\lambda = \hbar/mc$, as large as
a few microns. Hence, the limits on $\alpha$ may provide useful
guidance in narrowing down the myriad possible models linking
physics at very high energy scales to the much lower energy
Standard Model.

(c) Theories in which these hypothetical new forces arise from an
inverse-power potential, $V \propto r^{-p}$. An extensive review
of such inverse-power-law forces has been given recently in
Ref.~\cite{10d}.

Existing limits on such forces, when parameterized as in
Eq.~(\ref{1}), are relatively weak for several reasons. Most
significantly, if $\lambda$ is small then the effective
interacting masses are themselves necessarily small, and
background disturbances play a relatively more important role.
Secondly, for experiments with typical separations
$\stackrel{<}{\sim} 1\mu$m the dominant background arises from the
Casimir force \cite{review,l21}, which is not only relatively
strong over the relevant distances, but is also somewhat difficult
to completely characterize at the required level of precision
\cite{ours2}. Absent any alternative, limits on $\alpha =
\alpha(\lambda)$ have been inferred by developing detailed
theoretical models of the Casimir interaction \cite{ours2},
yielding the limit $\alpha \stackrel{<}{\sim} 10^{13}$ for
$\lambda \cong 150$ nm. Further improvements in experiments and in
our understanding of the Casimir force will lead to stronger
constraints on $\alpha(\lambda)$, but it is better to sidestep
theory altogether by carrying out what amounts to a
``Casimir-less" measurement.

In this Letter we report isoelectronic measurements (IET)
\cite{PLA,l22}, where the Casimir background is subtracted at the
outset, thus avoiding the necessity to model the Casimir force.
IET exploit the essentially electronic nature of the Casimir
force, whereas gravity and hypothetical forces couple to nucleons
and electrons. Hence, vacuum fluctuations cannot account for any
significant difference in the forces between a probe and two
materials with identical electronic properties.

A schematic of our set-up is shown in Fig. \ref{fig1}. We compare
the force {\em differences} over two dissimilar materials, Au and
Ge, which have been coated with a common layer of Au of thickness
$d_{Au}^{p} = 150 {\rm nm} > \lambda_p$, where $\lambda_p = 135$
nm is the plasma wavelength for Au. The fractional difference of
the Casimir force between two infinitely thick metallic plates and
two plates of thickness $d_{Au}^{p}$ is $\sim e^{-4\pi
d_{Au}^{p}/\lambda_{p}} \sim 10^{-6}$ \cite{[25a]}. In our
experiment, it translates to a difference $\Delta F_C \lsim
10^{-17}$\,N in the Casimir force between the Au coated sphere and
the two sides of the Au/Ge composite sample. Hence any
differential signal that the probe detects as it oscillates over
the underlying Au and Ge (which provide a large mass density
difference, $\rho_{Au} - \rho_{Ge} = 13.96 \times 10^3\,$kg/m$^3$)
must be due to an interaction via either gravity or some new
hypothetical force. Thus by directly comparing the forces on the
Au and Ge substrates a limit on $\alpha(\lambda)$ can be obtained,
without having to resort to a theory of the Casimir force for real
materials. The expression for this hypothetical force difference
is

\begin{eqnarray}
\Delta F^{hyp}(z)& = &-4\pi^2 G\alpha\lambda^3e^{-z/\lambda}R K_s  K_p, \label{intro1}\\
K_s& =
&\left[\rho_{Au}-\left(\rho_{Au}-\rho_{Cr}\right)e^{-d_{Au}^s/\lambda}
-\left(\rho_{Cr}-\rho_{s}\right)e^{-(d_{Au}^s+d_{Cr})/\lambda}
\right], \nonumber \\
K_p&=&\left[\left(\rho_{Au}-\rho_{Ge}\right)
e^{-\left(d_{Au}^{p}+d_{Pt}\right)/\lambda}
\left(1-e^{-d_{Ge}/\lambda}\right)\right], \nonumber
\end{eqnarray}

\nn where $G$ is the gravitational constant, $R \sim 50 \mu$m is
the radius of the sphere, $K_s$ ($K_p$) is a term associated only
with the layered structure of the sphere (plate), $\rho_s$,
$\rho_{Cr}$, $\rho_{Au}$, and $\rho_{Ge}$ are the densities of the
sapphire sphere, a Cr layer (used to increase Au adhesion to
sapphire), the Au layers and the Ge layer, respectively.
Thicknesses $d_i$ for the different materials are given in
Fig.~\ref{fig1}.

Since the hypothetical forces under study are weak \cite{ours2}, a
high sensitivity force measurement is required. A
microelectromechanical torsional oscillator (MTO) with a soft
spring $\kappa$ (high force resolution) and high quality factor
($Q$) satisfies the required demands. The MTO has a low coupling
with the environment \cite{ours2,ours1} $\kappa \sim
10^{-9}$\,Nm/rad, and $Q \sim 10^4$.

The films deposited on the sphere and the MTO  were characterized
by atomic force microscopy (AFM). A typical AFM line-cut at the
interface between the Au and Ge layers is shown in
Fig.~\ref{fig1}b. The observed ridge ( a valley in some samples)
arises from the imperfect alignment of the mask when depositing
the Au and Ge. The analysis of the AFM images indicates the
granular character of the samples, showing a maximum height
difference of 22\,nm. The average lateral dimension of the grains
was in the 100-150\,nm range, although grains as large as 500\,nm
were observed. The uncertainty in the position of the zero height
level (with respect to which the roughness is measured
\cite{ours2}) is $\sim 0.2$\,nm.

The experimental arrangement and calibrations performed are very
similar to those previously used to determine Casimir forces
\cite{ours1,ours2}. A voltage was applied to the sphere to
eliminate the residual electrostatic force caused by the
difference in work functions between the Au layer on the MTO and
the sphere. The angular displacement of the MTO, determined by
measuring the difference in capacitance between the MTO and the
underlying electrodes, $\Delta C = C_{right} - C_{left}$, yielded
the force acting on it. The sensitivity in the angular deviation
is $\delta \theta \simeq  10^{-9}{\rm rad}/{\sqrt{\mbox{Hz}}}$.

The force sensitivity is improved if the measurements are
performed at resonance, $\omega = \omega_o$. In this case, the
minimum detectable force is dominated by thermal fluctuations,
$\delta F(\omega_o) \simeq \delta F_{thermal} ={1}/{b}
\sqrt{{4\kappa k_B T}/({\omega_o Q}})$, where $k_B$ is Boltzmann's
constant. Consequently, it is necessary to measure the effect of
$\Delta F_{hyp}$ at resonance. $\Delta F_{hyp}$ can be described
as the product of a function of the separation $e^{-z/\lambda}$
and a function of the $x-$coordinate, the difference in the
average mass density. Hence, we induced a vertical oscillation on
the MTO such that the separation between the MTO and the sphere
changed as $z_m = z_{mo}+\delta z \cos(\omega_z t)$, with $z_{mo}
\gg \delta z$. We simultaneously moved the MTO along a direction
parallel to its axis, such that the effective mass density under
the sphere was $\rho_{eff} = \rho^+ + \rho^- \Xi(t)$, where $
\rho^{\pm}  = (\rho_{Au} \pm\rho_{Ge})/2$, and $\Xi(t)$ is a
square-wave function with characteristic angular period $T_x
=2\pi\omega_x^{-1}$. At $t = 0$ the sphere is positioned over the
MTO on the Au/Au half. The Casimir interaction leads to a shift of
the resonance frequency of the MTO from its natural oscillation
frequency $f_o$ to $f_r$ \cite{acl}. By selecting $\omega_z
+\omega_x = \omega_r$, $\Delta F_{hyp}$ has a Fourier component at
$\omega_r$ given by

\beq {\cal F}_{hyp} (z_{mo},\omega_r) = \Delta
F_{hyp}(z_{mo})\times \frac{2}{\pi}\times I_1\left (\frac{\delta
z}{\lambda}\right ), \label{eqforce}\eeq

\nn where $I_1$ is a Bessel function of the second kind. ${\cal
F}_{hyp} (\omega_r)$ is the only Fourier component with a
significant signal-to-noise ratio, even though no parts in the
system are moving at $f_r$. We selected $f_x = \omega_x/2\pi =
\omega_r/140 \pi=f_r/70\sim 10$\,Hz. Consequently, $f_z =
\omega_z/2\pi = 69 f_x$. The phases of the different signals were
chosen to simultaneously cross through 0 every $t_{xr} \equiv
70/f_r$ ($t_{xz} \equiv 69/f_z$) for the signals at $f_x$ and
$f_r$ ($f_z$). The synthesized signal at $f_r$ was used as
reference, see Fig.~\ref{exp1}a. The amplitudes were adjusted to
provide a peak-to-peak lateral displacement of $D \simeq 150 \mu$m
and a vertical amplitude that ranged between $\delta z \simeq$
10\,nm at the closest separations to $\delta z \simeq$ 50\,nm at
$z_{mo} = 500$\,nm.

For each resonant period of oscillation of the MTO, $T_r = 1/f_r$,
ten equally spaced data points ${\cal F}_d(t_{i})$ were acquired,
$t_{i} = i T_r/10$, $i = 1,\cdots,10$. Simultaneously, ${\cal
P}(t_{i})= {\cal F}_d(t_{i})\cos\omega_rt_{i}$ and ${\cal
Q}(t_{i}) = {\cal F}_d(t_{i})\sin\omega_rt_{i}$ were determined.
Averaging of ${\cal F}_d$, ${\cal P}$, and ${\cal Q}$ was achieved
by adding the signals for all different $T_r$ in the corresponding
$i = 1,\cdots,10$ intervals. Furthermore, the summation over $t_i$
of ${\cal P}(t_{i})$ (${\cal Q}(t_{i})$) yields the in-phase
${\cal F}_p$ (quadrature ${\cal F}_q$) Fourier component at
$\omega_r$.

A strong reduction of the random noise was achieved by increasing
the integration times $\tau$. $\tau$ was changed  between 0.1 and
2000 s in a 1, 2, 5, 10, $\cdots$ sequence. Fig. \ref{exp1} shows
the results obtained at a separation $z_{mo} = 300$ nm. Figs.
\ref{exp1}b-d show the observed behavior of ${\cal F}_p$ and
${\cal F}_q$ in one sample for three characteristic values of
$\tau$ (the number of repetitions $N$ decreased as $\tau$
increased). Fig. \ref{exp1}d, shows the behavior at $\tau =
2,000$\,s for all seven samples investigated. Several features
should be noted in Fig. \ref{exp1}: {\em (i)} for each sample and
all values of $\tau$ $|\overline {\cal F}_T| =
\sqrt{\overline{\cal F}_p^2 + \overline{\cal F}_q^2}$ is constant
within the statistical error. $\overline{\cal F}_p$
($\overline{\cal F}_q$) is the average of ${\cal F}_p$ (${\cal
F}_q$) over the different repetitions; {\em (ii)} $|\overline{\cal
F}_T|$ remains constant within a factor of 2 for different
samples; {\em (iii)} the phase $\Theta = \arctan(\overline{\cal
F}_q/\overline{\cal F}_p)$, however, assumes values close to
either 0 or $\pi$, indicating that the force is larger over either
the Au or the Ge side of the composite sample, respectively. We
assume that the forces measured correspond to either $\Theta = 0$
or $\pi$ \cite{reference}.

We calculated the force $\overline{\cal F}(\tau,z_{mo}) =
\overline {\cal F}_T(\tau, z_{mo}) \pm s_{N^*}(\tau, z_{mo})
t_{\beta,N^*}$ required to exclude, at the $\beta$ = 95 \%
confidence level, any hypothetical force of the form given by Eq.
(\ref{intro1}). Here $s_{N^*}(\tau, z_{mo})$ is the mean square
error of $\overline {\cal F}_T(\tau,z_{mo})$, $t_{\beta,N^*}$ is
the Student's coefficient, and  $N^* = 6$ is the number of
repetitions at $\tau = 2000$\,s. The + (-) sign is is used for the
in phase ($\pi$ out of phase) cases. Fig. \ref{exp3} shows the
asymptotic behavior ${\overline{\cal F}}(\tau =2,000$\,s$,z_{mo})$
for all samples. $|{\overline{\cal F}}(\tau =2,000$\,s$,z_{mo})|$
decreases as $z_{mo}$ increases, but not exponentially, as
expected from Eq. (\ref{eqforce}). Independently of the origin of
${\overline{\cal F}}(\tau =2,000$\,s$,z_{mo})$, however, the
shaded region in Fig. \ref{exp3} represents values of hypothetical
forces {\it excluded} by our experiment.

Although ${\overline{\cal F}}(\tau =2,000$\,s$,z_{mo}) \neq 0$, we
do not believe it originates from new physics. As noted above, its
$z_{mo}$ dependence is not exponential. More importantly, it
changes sign depending on the sample under study, showing no
correlation with the underlying Au and Ge layers. We therefore
attribute its presence to the manifestation in the Casimir force
of a change $\delta z' \sim 0.1$\,nm in $z_{mo}$, i.e. $z_{mo}$
takes different values over the Au and Ge sides. With the AFM
characterization, the height difference between the two sides is
not known to better than $\delta z' \sim$ 0.1\,nm
\cite{curvature}. This translates into a residual  Casimir force
difference $|\Delta F_C(200 \mbox{nm})| \sim 15$ fN, sufficient to
explain the observed $|\overline{\cal F}(\tau,z_{mo})|$.

We can, however, rule out other sources for the observed
background: {\it (i)} A motion of the sphere in a direction not
parallel to the MTO's axis was ruled out by performing the
experiments without crossing the interface, and at different
$z_{mo}$. The uncertainty $\delta y \sim$ 3\,nm in the motion of
the sphere over the $D = 150$\,nm excursion yields a $\sim$
0.2\,fN error. {\it (ii)} Local differences in roughness yield an
estimate for the residual force to be $\lsim 2$\,fN, while patch
potentials \cite{Umar} give a contribution $\lsim 5$\,fN. The
effect of roughness and patch potentials can be further reduced by
moving the sphere back and forth across the interface to points
{\it randomly} selected, with dispersion large enough to average
the local changes. This approach yielded the same results. {\it
(iii)} The effect of the ridge at the interface is also small. A
Fourier analysis of its contribution yields $\Phi (200
\mbox{\,nm}) \lsim 3$\,fN. As $z_{mo}$ increases, this
contribution decreases rapidly, having a value $\Phi(500
\mbox{\,nm}) \lsim 0.1$ fN. We note that for samples with a valley
instead of a ridge the contribution is negligible. {\it (iv)} We
also checked that the finite size of the sample does not affect
the Casimir background or the Yukawa corrections. {\it (v)}
Magnetic or gravitational (Newtonian) forces do not give a
measurable background at $\omega_r$. The component of the magnetic
force with a signal at $\omega_r$ is associated with the
preferential presence of magnetic impurities in one of the sides
of the engineered sample (either Ge or Au). This force, comparable
to the magnetic interaction between isolated atoms, is much
smaller than the sensitivity of our apparatus. The Newtonian
gravitational attraction difference between the sphere and the
composite Au/Ge sample is $\sim 3 \times 10^{-21}$\,N, independent
of separation. Hence, this force not only is too small to be
detected, but it also does not provide a Fourier component at
$\omega_r$.

The results shown in Fig. 3 can be used to obtain more stringent
limits on hypothetical forces since these must be less than, or at
most equal to, the observed background. The values of forces in
the shaded region in Fig. \ref{exp3} have been experimentally
excluded. Furthermore, the $\Theta = 0$ and $\Theta = \pi$ cases
yield very similar absolute values for ${\overline{\cal F}}$.
Hence, for a given $z_{mo}$, we use the smallest $|{\overline{\cal
F}}(\tau =2,000$\,s$, z_{mo})|$ as the maximum allowed
hypothetical force. Using this value in the left-hand side of
Eq.~(\ref{eqforce}) we obtain an $\alpha (\lambda)$ curve.
Repeating this procedure for different $z_{mo}$ we obtain a family
of curves whose envelope provides the strictest limits arising
from our experiments. This curve, together with previous results,
is shown in Fig. \ref{cons1}, which also shows the regions in
{$\lambda,\alpha$} phase space where different models predict the
existence of new forces \cite{limits}. Our realization of a
``Casimir-less" IET yields a $\sim$ 10-fold improvement (in the
[30, 400] nm range) on existing limits for Yukawa-like corrections
to Newtonian gravity. This has significant consequences for models
of moduli exchange, as proposed by supersymmetry, by further
constraining the supersymetric parameters. We believe that our
direct, improved experimental test at submicron separations will
continue to motivate theoretical development for this range of
separations. We also note that our experiment can be improved by
gluing the sphere to the MTO and oscillating the test masses over
it. By judiciously designing the test masses, most of the problems
associated with the background can be removed. In this scenario,
limits on $\alpha$ down to 10$^{6}$ can be achieved at separations
$z_{mo} \sim 100$\,nm.

We are deeply indebted to U. Mohideen for suggesting the
randomization of the motion across the interface and for
stimulating discussions. R.S.D. acknowledges financial support
from the Petroleum Research Foundation through ACS-PRF No.
37542--G. The work of E.F. is supported in part by the U.S.
Department of Energy under Contract No. DE--AC02--76ER071428.


\newpage
\begin{figure}
\caption{\label{fig1} {\bf (a)} Scanning electron microscope image
of the MTO with the composite sample deposited on it. Inset:
Schematic of the sample deposited on the MTO. The coordinate
system used in the paper is indicated. {\bf (b)} Diagram of the
experimental set-up. Inset: AFM profile of the sample across the
interface. $d_{Cr} = 1$\,nm; $d_{Au}^s =200$\,nm: Au layer
thickness on the sapphire probe; $d_{Ge} = 200$\,nm: thickness of
both the Ge (indicated in orange) and bottom Au layers on the
substrate; $d_{Pt} = 1$\,nm: thickness of the Pt film (used to
avoid diffusion of Ge in Au). There is also a $d_{Ti} = 1$\,nm
layer of Ti deposited to increase adhesion to the MTO. }
\end{figure}
\begin{figure}
\caption{\label{exp1} {\bf (a)} Integration time, $\tau$,
dependence of the signal. Data at the shortest integration time
have been scaled by $A$ = 20. Data were obtained at $z_{mo} =
300$\,nm. The dotted line is the measured reference signal at
$\omega_r$, adjusted to fit on the same scale. {\bf (b)} through
{\bf (d)} Measured ${\cal F}_{p}$ and
 ${\cal F}_{q}$ signals with $\tau$ = 2\,s ($N = 300$), 200\,s ($N = 30$), and
2,000\,s ($N = 6$), respectively. Each point is a repetition of
the experiment. The circles in {\bf (b)} and {\bf (c)} are
centered at (${\overline{\cal F}}_p$,${\overline{\cal F}}_q$) and
have a radius given by the calculated thermal noise over the
relevant $\tau$. The measured noise, obtained at large $z_{mo}$,
exceeds the thermal noise by $\sim$ 20\,\%. {\bf (d)} Results at
$\tau$ = 2000\,s for all samples.}
\end{figure}
\begin{figure}
\caption{\label{exp3} Dependence of the peak-to-peak Fourier
component ${\overline{\cal F}}_T(\tau =2,000$\,s$,z_{mo})$ on
$z_{mo}$ for all measured samples. The shaded region in the figure
represents values of hypothetical forces excluded by our
experiments. Colors and symbols for the different samples are the
same as those used in Fig.~\ref{exp1}d. The upper (lower) hatched
area is defined by the points with minimum absolute value of
$\overline {\cal F} (\tau,z_{mo}) = \overline {\cal F}_T(\tau,
z_{mo}) + s_{N^*}(\tau, z_{mo}) t_{\beta,N^*}$ ($\overline {\cal
F} (\tau,z_{mo}) = \overline {\cal F}_T(\tau, z_{mo}) -
s_{N^*}(\tau, z_{mo}) t_{\beta,N^*}$).}
\end{figure}
\begin{figure}
\caption{\label{cons1} Values in the {$\lambda,\alpha$} phase
space excluded by the experiment. The red curve represents the
limit obtained in the current experiment. Curves 1 to 5 were
obtained by Mohideen's group \cite{Experimental2}, our group
\cite{ours2}, Lamoreaux \cite{Experimental1}, Kapitulnik's group
\cite{Experimental3}, and Price's group \cite{Long},
respectively.}
\end{figure}


\begin{thebibliography}{99}
\bibitem{acl} $f_r$ is a function of the separation due to the
non-linear dependence of the Casimir force \cite{ours2,ours1}.
\bibitem{reference} The deviations of $\Theta$ with respect to 0 and $\pi$ are
comparable to those observed in the system when the experiment is
replaced by a frequency mixer. This leads us to believe that these
deviations are artificial and induced by the electronic circuitry.
\bibitem{curvature} $\delta z'$ is not expected to be originated in
curvature differences between the MTO and the platform. These
differences, along the direction of motion of the sphere, are
smaller than 0.1\,nm, the measurement error (WYCO NT3300 optical
profilometer). The sample dependant curvatures are a few nm over
the 500\,$\mu$m of the MTO's length.
\end{thebibliography}
\end{document}